\UseRawInputEncoding

\documentclass[twocolumn]{aastex701}
\usepackage{booktabs}

\begin{document}

\title{SoFTg: A hybrid segmentation framework for solar granulation revealing turbulent convection statistics at DKIST scales}

\correspondingauthor{Michele Berretti}
\email{michele.berretti@unitn.it}

\author[0009-0007-2465-1931]{M. Berretti}
\email{michele.berretti@unitn.it}
\affiliation{University of Trento \\
Via Calepina 14, 38122 Trento, Italy}
\affiliation{University of Rome Tor Vergata, Department of Physics \\
Via della Ricerca Scientifica 3, 00133 Rome, Italy}

\author[0000-0002-4525-9038]{S. Criscuoli} 
\email{}
\affiliation{National Solar Observatory, 3665 Discovery Dr., Boulder, CO 80303, USA}

\author[0009-0003-0764-3965]{A. Fanti}
\email{}
\affiliation{Department of Computer, Automation and Management Engineering, Sapienza University of Rome, via Ariosto 25, 00185, Roma, Italy}

\author[0000-0001-8556-470X]{P. H. Keys}
\email{}
\affiliation{Astrophysics Research Centre, School of Mathematics and Physics, Queen’s University Belfast, Belfast, BT7 1NN, Northern Ireland, UK}

\author[0000-0002-1155-7141]{L. Bertello}
\email{}
\affiliation{National Solar Observatory, 3665 Discovery Dr., Boulder, CO 80303, USA}

\author[0000-0002-5365-7546]{M. Stangalini}
\email{}
\affiliation{ASI Italian Space Agency \\
Via del Politecnico snc, 00133 Rome, Italy}

\author[0000-0002-2276-3733]{F. Berrilli}
\email{}
\affiliation{University of Rome Tor Vergata, Department of Physics \\
Via della Ricerca Scientifica 3, 00133 Rome, Italy}

\begin{abstract}

In this work, we present Solar Feature Tracking: granules (SoFTg), a new hybrid algorithm within the SoFT library designed for the segmentation and tracking of solar granulation. By combining a heuristic approach based on a custom adaptive-thresholding technique with the speed and robustness of deep-learning methods, SoFTg provides a fast, reliable and user-friendly solution for granule segmentation, followed by the standard SoFT approach for tracking. We present a comprehensive evaluation of the algorithm’s performance using multiple metrics and analyse the different stages of the processing pipeline. The method is tested on three datasets acquired with the Visible Broadband Imager operating in the red continuum at the Daniel K. Inouye Solar Telescope in Maui, Hawaii. We present distributions of both the lifetime and equivalent diameter of the thousands of granules tracked, providing, for the first time, statistically robust evidence of turbulent convection at the spatial scales now accessible through the high-resolution imaging capabilities of DKIST.

\end{abstract}



\section{Introduction} 

\begin{figure*}[ht!]
    \centering
    \includegraphics[]{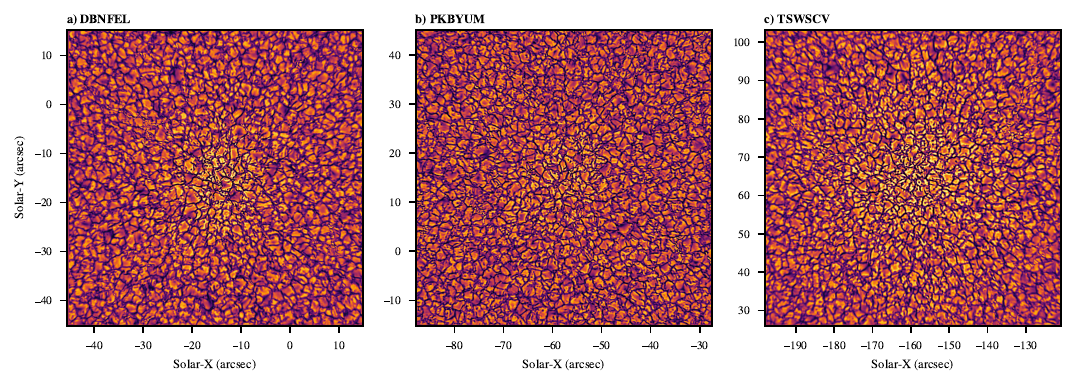}
    \caption{{Sample field of view of the three datasets used in this work after pre-processing.} The images have been enhanced with the pre-processing pipeline described in the Section~\ref{subsec:prepro} in order to increase the sharpness and contrast, leading to improved detection of granules. The title of each panel reports the dataset ID of the DKIST data available on the publicly accessible data portal. The first frame in the dataset was used for all three panels.}
    \label{fig:1}
\end{figure*}

First observed by W. Herschel in 1801 \citep{herschel_observations_1801}, solar and stellar granulation is the primary visual manifestation of convective energy transport in stellar atmospheres, playing a key role in organising magnetic fields and driving photospheric dynamics. Over the past decades, numerical and observational work has significantly reshaped our understanding of this phenomenon, although the precise physical mechanism of granulation and the link between small-scale behaviour and global-scale dynamics remain largely unknown. While classical theories framed granules as convective cells overshooting into the photosphere \citep{bohm-vitense_uber_1958, Nordlund_1977}, simulations carried out by \citet{stein_simulations_1998} and \citet{nordlund_solar_2009} established that the pattern is instead mainly driven by radiative cooling at the surface. In addition to simulations, observational studies \citep[e.g.][]{hirzberger_time_1997, Berrilli1999, Consolini1999, berrilli_2-d_2002} detailed the vertical dynamics of convective flows alongside the physical and geometric properties of individual granules. They demonstrated clear correlations between intensity variations, the underlying upflows and downflows, and the mechanisms driving reverse granulation and exploding granules. 

Moreover, this convective process also manifests itself at different spatial scales, as shown by \citet{Cattaneo2001,Roudier2003,Berrilli2004}. 
In particular, \citet{berrilli_spatial_2005} demonstrated that segmented photospheric granules, analyzed by means of a spatially resolved hexagonal  information entropy measure, can be used to investigate the space–-time formation and clustering of photospheric structures across scales ranging from granulation to mesogranulation.

Furthermore, tracking the horizontal evolution of individual granules allows us to use them as proxies of large-scale velocity fields, as shown by several authors \citep[e.g.][]{november_precise_1988,brandt_vortex_1988,rutten_details_1989,rieutord_granules_2001,potts_balltracking_2004,rieutord_solar_2008,roudier_large-scale_2018}. This horizontal advective motion plays a fundamental role in organising magnetic field concentrations across various spatial scales in the photosphere \citep{spruit_solar_1990}. These magnetic fields are brought to the surface by convective upflows and are subsequently swept towards the boundaries of granular cells \citep{stenflo_magnetic-field_1973, jess_study_2010, riethmuller_comparison_2014, borrero_solar_2017, roberts_mhd_2019}. Furthermore, the interplay between moving granules and the magnetic field is thought to be the origin of vortex-like flows in the solar photosphere \citep[see][]{Tziotziou2023}. Therefore, being able to confidently detect and study granules is a major task which requires sophisticated image segmentation techniques capable of discerning complex-shaped blobs with dynamical boundaries, a prerequisite for a deeper understanding of convective energy transport mechanisms within the envelopes and atmospheres of the Sun and other solar-type stars.

Historically, many different approaches have been tested by different authors. Although a thorough review of all the segmentation algorithms ever realised for solar granulation is beyond the scope of this work, we will briefly discuss a few of the most widely used techniques here. In \cite{roudier_structure_1986}, the authors introduced a Fourier-based segmentation technique, later reprised by \cite{hirzberger_time_1997}, with filter constants tuned to match intensity inflection points determined via a Laplacian operator. Applied exclusively to reconstructed broad band images, the technique divides intergranular lanes between adjacent granules to define granular cells. In more detail, to identify individual granules, their segmentation technique defines an intensity threshold level at $1.01\bar{I}$ (where $\bar{I}$ is the mean intensity of the measured granulation field). The algorithm then views any contours enclosing values above this threshold as granules prior to removing features smaller than a cut-off area of $4\text{ pixels}$ to reduce noise.\\
In the paper by \cite{Leitzinger2005}, segmentation of mesogranules is performed by applying a threshold method to divergence maps derived from photospheric flow fields. Pixels with divergence values above a given cumulative histogram level (80\%, 85\%, 90\%, or 95\%) are defined as mesogranular regions, creating binary segmented maps.\\
Departing from the standard spectral-based methods of the time, \cite{Berrilli1999} employed an adaptive thresholding algorithm to segment granules and a skeletonizing algorithm, based on an iterative Medial Axis Transformation (\textit{i}-MAT algorithm), to segment convective structures by combining velocity and intensity fields derived from photospheric lines probing different atmospheric heights. This segmentation allowed the statistical properties of upflow and downflow regions to be characterized as a function of height, revealed pronounced sign-singularity scaling of the vertical velocity field from scales of tens of megameters down to the granular scale, and showed that a few energetically dominant modes suffice to reconstruct both the convective field and the five-minute oscillation field \citep{Berrilli1999,Consolini1999,Carbone2002,berrilli_2-d_2002}. This  numerical approach was further developed in \cite{berrilli_spatial_2005}. A third method commonly used exploits different intensity clipping thresholds to ensure optimal adaptation to the solar structures under study. This method is known as `multi-level tracking' and was introduced to the study of solar granulation in \cite{bovelet_new_2001}. The algorithm returns individual sizes in pixel counts and is adapted to represent inter-granular structures realistically. A similar approach was tested by \cite{moro_solar_2004} and is referred to as `Two-level Structure Tracking' (TST). More recently, \cite{falco_comparison_2017} introduced a new custom technique based on local maxima detection, slope analysis, and intensity thresholding.\\
In the paper by \cite{Lemmerer2014}, segmentation of granules is performed using a two-dimensional algorithm applied to high resolution solar radiation hydrodynamics simulation data. The method combines multiple-level thresholding on vertical velocity profiles with image-processing techniques, including Laplacian-of-Gaussian edge detection and morphological operations, applied to emerging intensity maps on the optical depth unity surface. 
More recently, \cite{Liu2021} developed a  granule segmentation method based on edge detection, extracting pixels with local intensity minima as edges. False edges are removed using intensity thresholds and size criteria. Morphological operations (erosion and dilation) separate closely connected granules. Finally, magnetic bright points are identified and removed, yielding pure granule cells across full length scales.

However, while all of these techniques have provided valuable insight into the nature of solar granulation, they also present limitations that cannot be overlooked. These include a strong dependence on the adopted intensity threshold, limited adaptability to variations in image contrast, the need for several empirical decision criteria, and sensitivity to internal intensity minima and maxima, which can lead to the fragmentation or merging of individual structures. Moreover, \citet{moro_solar_2004} highlighted the difficulties involved in directly comparing the performance of different segmentation approaches across datasets with different observational characteristics. As we enter an era of increasingly large and high-quality solar datasets, these limitations become particularly relevant: a modern segmentation technique should not only provide reliable identification of individual granules, but also do so rapidly and consistently across different observing conditions, while preserving the temporal coherence required for subsequent feature tracking and velocity-field reconstruction.

To this scope, \cite{armstrong_fast_2019} and \cite{AsensioRamos2023} have highlighted the potential that machine learning approaches, and in particular deep learning, hold to greatly improve our capabilities of processing the ever increasing volume of data from observations and simulations. Specifically for the case of granules, in \cite{diaz_castillo_towards_2022}, the authors developed a new deep-learning approach to solar granulation segmentation that can automatically segment and classify granules in images from the Imaging Magnetograph eXperiment \citep[IMaX]{MartinezPillet2011}. This approach is based on a fully convolutional U-Net neural network \citep{ronneberger_u-net_2015} and can simultaneously classify the detected granules into different categories (complex-shaped, regular, dark and exploding). Further efforts have attempted to adapt this model to other high-resolution datasets; for example, \citet{mansouri_exploring_2025} used it with observations from the 4-m Daniel K. Inouye Solar Telescope \citep[DKIST, ][]{rimmele_daniel_2020}.

n this work, we present SoFTg, a novel granulation segmentation technique that prioritises ease of use, adaptability, and multi-instrument applicability, enabled by its instrument-independent design rather than by reliance on instrument-specific features. By combining classic methods with modern machine learning approaches, our aim is to develop a fast, easy to use and adaptable granule segmentation routine. Thanks to our custom granule labeller, which is based on an adaptive threshold and watershed algorithm, we can easily create a large training and validation dataset with little or no user input while retaining control over the data on which the model learns. This approach enables users to apply our method to virtually any dataset, overcoming the instrument constraints of previous approaches. Here, we demonstrate its performance using high-resolution observations of three quiet Sun patches taken with the Visible Broadband Imager \citep[VBI][]{woger_daniel_2021} mounted at DKIST, while providing a thorough analysis of the method and model. Finally, we extract and analyze the lifetime and equivalent diameter distributions of nearly twenty thousand tracked granules, providing the first statistically robust observational sample of turbulent convection at DKIST spatial scales.

\begin{figure*}[ht!]
    \centering
    \includegraphics[width=\linewidth]{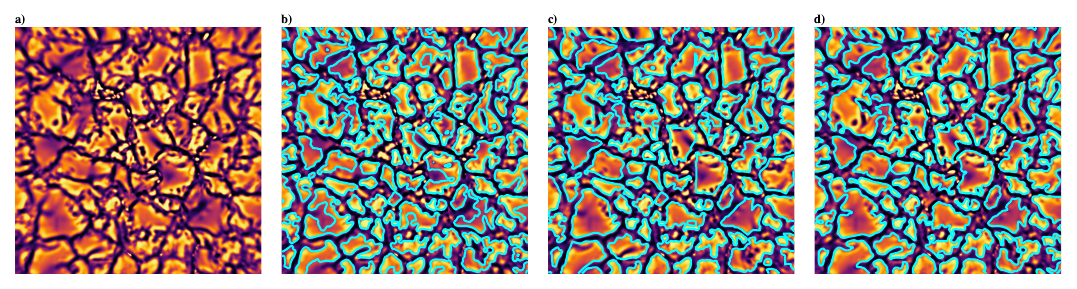}
    \caption{\textbf{a.)} Sample crop of the FOV of the DBNFEL dataset.  \textbf{b.)} Example of the performance of the adaptive threshold step on a sample DBNFEL dataset frame as described in Sect.~\ref{subsec:adaptive}. \textbf{c.)} Ground truth for the same frame obtained by manually refining the detected features from the adaptive threshold step, as shown in Sect.~\ref{subsec:manual}. \textbf{d.)} Final prediction from the UNET++ model, see Sect.~\ref{subsec:unet}. In all three panels, the cyan lines highlight the boundaries of the detected granules.}
    \label{fig:2}
\end{figure*}

\section{Dataset}

To train and showcase the results of our code, we selected three different high-resolution quiet Sun patches near the centre disk observed with the same VBI red channel in the red continuum (Ni~{\sc i} $668.4\pm0.4$~nm). The instrument has a spatial sampling of $0{\,}.{\!\!}{''}017$~pixel$^{-1}$ and observes a $69''\times 69''$ field of view with a typical temporal cadence of $33$~ms. Burst of 80 frames are used for the reconstruction, so that the resulting temporal cadence of each dataset is 2.667 seconds. Although these parameters may vary slightly depending on the observation campaign conditions, for our work, we ensured only selecting a dataset with optimal seeing conditions. The three datasets used in this work are characterised as follows.

\begin{itemize}
    \item DBNFEL, L1-ZUBLA: Part of an observational campaign to study the dynamics of bright-points in the photosphere, it consists of a 50 minutes time sequence of a quiet Sun patch positioned at $(-45.55, -45.19)''$, $(14.88, 15.24)''$ from disc centre. The average seeing Fried parameter $r_0$, which quantitatively expresses the image degradation due to atmospheric turbulence, and with larger values meaning weaker turbulence and thus better seeing, is $0.08\pm0.03$~m. 
    \item PKBYUM, L1-DNZOE: Data acquired to study small-scale flows of turbulent convection, it consists of a 50 minute time sequence of a quiet Sun patch positioned at  $(-87.83, -15.21)''$, $(-27.46, 45.15)''$ from disc centre with an average Fried parameter $r_0=0.07\pm0.02$~m. 
    \item TSWSCV, L1-OEASK: Data acquired to study scattering polarisation, it contains images of a quiet Sun region at  $(-198.03, 25.98)''$, $(-120.78, 103.21)''$ from the centre disc. The sequence has a duration of 35 minutes with an average Fried parameter $r_0=0.06\pm0.02$~m. 

\end{itemize}

The three datasets are publicly available at the DKIST data centre. All images were reduced using the standard VBI reduction pipeline, and speckle reconstructed \citep{woger_speckle_2008} to compensate for residual image degradation due to seeing. A sample of the three field of view (FoVs) is shown in panels a), b), and c) of Fig.~\ref{fig:1}, respectively.

\section{Methods}\label{sect:meths}

This novel suite for segmenting granules can be divided into three distinct steps, which must be followed to ensure optimal segmentation results, that is, the initial labelling process, followed by manual refinement, and finally, the training of the UNET++ model. While a pre-trained model is provided for convenience, its peak performance is achieved exclusively with VBI red continuum images, as these were used for training. Here, we will thoroughly discuss all three processes. 

\subsection{Pre-processing pipeline}\label{subsec:prepro}
A pre-processing pipeline is applied to optimise image contrast and sharpness for accurate segmentation. This step begins with a temporal median stack of contiguous frames $N$. The normalised median image is then processed using Contrast Limited Adaptive Histogram Equalisation (CLAHE) to increase the local feature contrast. Subsequently, an unsharp mask is applied by subtracting a blurred background to emphasise structural edges. Because the natural dynamics of solar granulation can cause motion blur if too many frames are aggregated, the stacking window must remain well below the granular evolutionary timescales. Given the 2.67~s cadence of our datasets, we set $N=5$, creating an effective integration window of $\sim$13~s. This configuration successfully mitigates high-frequency atmospheric noise without degrading the physical boundaries of the granules. The impact of this pre-processing pipeline is illustrated in Figure~\ref{fig:ap1} for a sample frame of the DBNFEL dataset.

\subsection{Adaptive thresholding}\label{subsec:adaptive}

In order to simplify the labelling process, allowing larger quantities of files to be processed, increasing the training set, and ensuring overall better performances, we developed a custom adaptive threshold and watershed segmentation algorithm to provide an initial estimate of the labels for each granule. Based on observational properties of the dataset, such as the diffraction limited angular resolution and the detector spatial sampling of the instrument, as well as physical properties of granulation, as the expected size for both granules and intergranular lanes, we apply a simple threshold discriminator on a rolling window shaped as a Gaussian (considered a zero-th order approximation of the PSF of any instrument) with a threshold based on the intensity levels of the portion of the FoV as follows:

\begin{equation}
    \mathcal{T} = \tilde{I} - k \sqrt{\mu_{\text{sq}} - \mu^2}
\end{equation}
where $\mathcal{T}$ is the local threshold, $\tilde{I}$ is the local median, $\sqrt{\mu_{\text{sq}} - \mu^2}$ is the local standard deviation while $k$, being the only free-parameter in this step, consists of a relaxation factor for the threshold and should be adjusted depending on the dataset. For our three datasets, we used a $k$ value of 0.4 and estimated that the size of the granules is 1250~km and the intergranular lanes are 150~km. It is important to note that these parameters are subject to variation between different datasets. Previous studies have demonstrated that the distributions for these features can be quite broad, making it overly simplistic to rely on a single fixed parameter \citep{1989ApJ...336..475T, hirzberger_time_1997, 1998ASPC..140..455R, 2002A&A...383..275H}.

Once the initial binary mask is produced, the intergranular lanes are cleaned through a series of erosions and dilations with a disk-shaped footprint the size of the input expected size for lanes. Smaller features can also be selectively rejected as granules using a user-defined input parameter that sets the minimum detection size, here it was set to one tenth the typical size of granules known in literature (i.e. $125$~km). Then, using morphological skeletonisation, it estimates the skeleton of intergranular lanes and prunes dead ends. This should effectively improve the detection of convex-shaped granules resulting from a colder core, which is typical of exploding or exploded granules \citep{hirzberger_time_1999}. Finally, following the standard procedures for watershedding, we estimate the basin point with a Euclidean Distance Transform and proceed to flood with watershed to provide the final preliminary labels, ready for either manual refinement or direct training of the UNET++ model. An example of the initial labels is shown in panel a of Figure~\ref{fig:2}. 

\begin{figure*}[ht!]
    \centering
    \includegraphics[width=\linewidth]{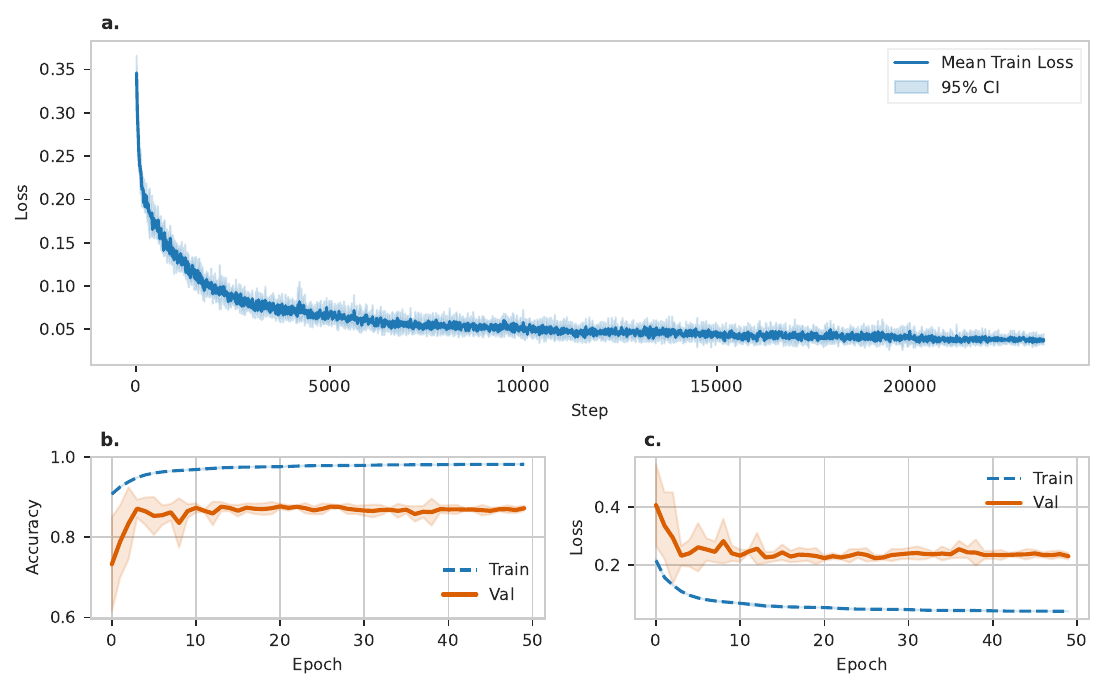}
    \caption{Overview of the training process diagnostics. \textbf{a)} Mean training loss as a function of the training step (solid line), bounded by a shaded 95\% confidence interval (shaded region) aggregated across all seed runs. \textbf{b)} Train (dashed line) and validation (solid line) epoch-averaged accuracies, presenting the mean trajectory across all runs bounded by its 95\% confidence interval. \textbf{c)} Train (dashed line) and validation (solid line) epoch-averaged loss across all runs with corresponding shaded 95\% confidence intervals. A unified color scheme is applied across all subplots to clearly distinguish training metrics (blue) from validation metrics (orange).}
    \label{fig:3}
\end{figure*}

\subsection{Manual label refinement}\label{subsec:manual}

Once the initial labels are prepared, they must be manually checked and fixed to ensure the best possible results for segmentation. To achieve this, we developed a streamlined label editor that allows users to simply drag the borders of pre-detected granules, remove them entirely, or add new ones. The process was streamlined in order to allow for easy processing of many frames allowing for larger and ad-hoc training sets. In our case, a single annotator curated 27 distinct masks in approximately 4 hours, with all final labels reviewed and approved by consensus among all authors.

\subsection{UNET++}\label{subsec:unet}

We believe that explaining the architecture of the model in its full details would be beyond the scope of this work. In this section, we will explain the reasons why we opted for a deep learning approach to refine our initial labels, why it works well, and finally a brief description of the model we used. Deep learning models offer a clear speed advantage over classical segmentation techniques. Given the massive data volumes typical of modern solar physics, adaptable and high-throughput computational workflows are essential. While deep learning algorithms are frequently critiqued as ``black boxes'' due to their uninterpretable internal mechanics, this architecture can be strategically isolated within a pipeline where internal transparency is secondary to empirical performance. In this framework, the neural network functions strictly as an optimised inference engine trained to generalise labels across a homogeneous dataset. This accelerates the feature-extraction pipeline while retaining full user control over both the input parameters and the downstream analysis, thereby bounding the model's operational freedom. In contrast, relying solely on an adaptive threshold proves insufficient; it struggles to resolve smaller granulation structures, and its computational footprint scales poorly. Integrating an adaptive window enhancement would further exacerbate processing times, rendering classical approaches impractical for large-scale, multi-year solar datasets.

Finally, following label refinement, we can train our UNET++ model, built on the standard UNET model with the addition of dense skip connections \citep{zhou_unet_2018}. UNET (and its upgraded variant) is a classic deep learning model, first introduced in the medical field \citep{ronneberger_u-net_2015} where precisely recognising the boundaries of features is a critical lifesaving task. The standard UNET model is a convolutional neural network optimised to provide precise segmentation maps with fewer training images than other models would require. While most standard image-processing networks aim at scaling down the features in an image to simply understand what is observed (i.e. a picture of a cat), UNET adds a second layer that upsamples the images in order to not only understand what it is observed but also where it is placed in the high-resolution space of the original image. Hence, the U-shaped architecture and its name. Here, the dense convolution blocks added by UNET++ play their role, blending the features across multiple scales and providing semantic links between the high-resolution and low-resolution layers.

\begin{table*}[t]
\centering
\caption{Comparison of the segmented granular pattern with the ground
truth. The number of granules counts spatially connected granular
regions. Intergranular lanes are reduced to one-pixel-wide central
lines to measure their structure. The centreline pixel count measures
their total extent; lane endpoints are pixels with only one neighbouring
centreline pixel, while branching pixels have at least three neighbours.
A single branching location may contain several branching pixels.
The endpoint density is the number of endpoints divided by the
centreline pixel count, with higher values indicating more free ends
per unit length. Percentages indicate relative differences from the
ground truth.}
\label{tab:topological_metrics}
\begin{tabular}{l c c c}
\toprule
\textbf{Metric} & \textbf{Ground Truth} & \textbf{Adaptive} & \textbf{UNet} \\
\midrule
Number of granules
    & 725 & 624 (-13.93\%) & 749 (+3.31\%) \\
Intergranular centreline pixels
    & 176212 & 186579 (+5.88\%) & 170054 (-3.49\%) \\
Lane endpoints
    & 2697 & 2762 (+2.41\%) & 2831 (+4.97\%) \\
Lane branching pixels
    & 4127 & 4747 (+15.02\%) & 4087 (-0.97\%) \\
Endpoint density (pixel$^{-1}$)
    & 0.03 & 0.02 (-33.33\%) & 0.03 (0.00\%) \\
\bottomrule
\end{tabular}
\end{table*}

In our approach, the segmentation task is deliberately formulated as a binary semantic segmentation problem, with only two classes: granules and intergranular lanes. The network therefore does not need to distinguish between different types of granular structures or identify individual granules as separate instances. Instead, its sole task is to assign each pixel to one of the two components of the photospheric granulation pattern. This substantially simplifies the learning problem and allows the network to focus specifically on accurately identifying the boundaries between granules and intergranular lanes. As a result, the model does not need to learn the additional complexity associated with multi-class or instance-level segmentation, making the training and subsequent inference considerably more straightforward.

In our setup, model training required approximately 36 hours on a four-GPU configuration with 128 GB of total VRAM. While this upfront computational cost is non-trivial, the resulting inference speed provides a substantial net gain in time efficiency when processing large datasets. Furthermore, because the U-Net architecture is relatively lightweight, it does not strictly require high-end computational clusters; it can also be trained on much more modest hardware. Memory requirements scale primarily with the training patch size and batch size, both of which can be tuned by the user to match local resource constraints.

\begin{figure*}[ht!]
    \centering
    \includegraphics[width=\linewidth]{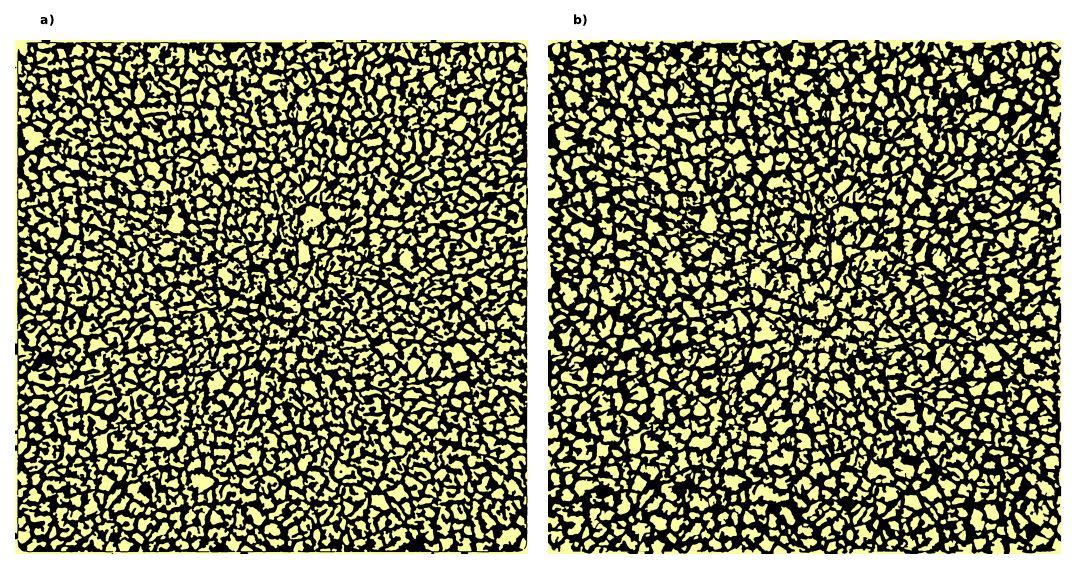}
    \caption{Comparison between the adaptive threshold estimated labels and the UNET++ final predictions. It can be clearly seen that the deep learning refinement greatly improves the initial guessing of the morphological properties of granulation, providing a much similar picture of what we would expect from solar granulation.}
    \label{fig:4}
\end{figure*}

\subsubsection{Data augmentation and loss function}

During data set construction, the images of the training and validation set are subject to geometric, spatial, and photometric augmentation. This process expands the training set without the need to provide additional hand-labelled data, improving the robustness of our training at virtually no cost \citep[see][]{wang_comprehensive_2024}. A summary of the different transformations can be found in Appendix A. For our analysis, we used only the first frame from the three datasets, divided into nine $1024\times1024$ patches. 

The model is trained using a differentiable Soft Intersection-over-Union (IoU) loss \citep{wang_jaccard_2023, jaccard_distribution_1912}, a region-based metric that rewards overlap between prediction and ground-truth instead of a per-pixel loss function. This choice was made because solar granule segmentation suffers from a class imbalance, where granule pixels significantly outnumber intergranular lane pixels in a single FoV. To further penalise the under-segmentation of granules, we upweighted the granule class by a factor of 2, corresponding to the inverse class distribution.

\subsection{Tracking and tabulation}

SoFTg is built on the SoFT library \citep{berretti_soft_2025}, inheriting many of its basic functionalities and optimising its core functions for granule segmentation and tracking. Although the detection and identification steps underwent the most important changes, with a completely new detection algorithm and machine learning implementation, the tracking and tabulation phases remain identical. Specifically, features continue to be associated across consecutive frames using a maximal spatial overlap criterion. In particular, this mutual overlap association is now executed sequentially rather than in parallel; this design choice mitigates the maximum RAM overhead when processing extensive sequences with high feature densities. Finally, the tabulation module functions identically to the original pipeline developed for magnetic features. For further details, we refer the reader to \cite{berretti_soft_2025}.

\section{Results \& Discussion}

In Figure~\ref{fig:2} we have shown the segmentation mask in the various steps of our approach. In the adaptive threshold step (panel a), the granules are detected correctly, and each seed actually isolates a different granule. However, the limitations of this approach are largely evident. Indeed, the boundaries of most exploding granules, characterised by a darker core, are not correctly segmented, creating convex-shaped granules that do not reflect the physical process behind them. Although standard overlapping pixel metrics are commonly used, they do not capture the true morphological properties of the considered network \citep{MaierHein2024}. Indeed, pixel counting metrics heavily reward volume overlap, which would result in a competitiveness of the adaptive threshold method, despite producing a visually broken segmentation. 

To properly quantify the visual and structural advantages of our hybrid approach, we evaluated the underlying topology of the segmented networks by extracting their morphological skeletons. Specifically, the number of connected components represents the discrete count of isolated foreground structures (i.e., individual granular cells). The total length of the network measures the cumulative pixel perimeter of the structural boundaries, while junctions define the nodes where three or more skeleton lines intersect, serving as a direct proxy for the interconnectivity of the network. Lastly, dead ends track unconnected line terminations, which are normalised against the total network length to yield a fragmentation index.

The network statistics shown in Table~\ref{tab:topological_metrics} reveal that the adaptive approach severely over-segments the background lanes, inflating the total network length to $116,880$~pixels and fabricating a massive surplus of false intersections ($4,747$ junctions compared to the ground truth of $4,127$). Furthermore, it merges or completely omits different granule bodies, underestimating the total number of connected components ($624$ compared to the ground truth of $725$). In contrast, the UNet prediction preserves the true network connectivity, closely mirroring the ground truth distribution across all topological indicators with $735$ components, $4,087$ junctions and an identical fragmentation index of $0.03$. Although dead ends are slightly more frequent in the UNet model ($2,762$), we wish to stress that they simply indicate where a thin granule boundary became too faint to be bridged by the network; the adaptive threshold method avoids this only by virtue of its aggressive over-segmentation. Therefore, the morphological properties of the networks highlight the ability of the UNet approach to accurately capture the observed granulation pattern, which standard pixel-based metrics fail to recognise.

Figure~\ref{fig:3} presents the training diagnostics of our model, statistically aggregated between all independent seed runs. In total, five different runs were carried out with each lasting about 36 hours. Panel a presents the mean training loss as a function of the training step, bounded by its 95\% confidence interval, which shows a steep initial descent during the early phases followed by convergence to a low-noise plateau. Panel b reports both the mean training and validation accuracies along with their corresponding variance bands, confirming the high efficacy and stability of the model across both phases. Finally, the epoch-averaged training and validation losses in panel c exhibit a consistent downward trajectory with narrow confidence intervals, demonstrating the model's capacity for generalisation and confirming successful learning without overfitting.

As previously emphasised, the core novelty of our approach lies in its adaptable framework, which can be readily applied to any high-resolution observation (or simulation) of solar granulation with minimal user input and modest computational resources. Consequently, providing a single pre-trained model would be contradictory to our primary objective. We stress that the diagnostics presented in Figure~\ref{fig:3} should be viewed as a representative example of a successful training run, serving as a benchmark for our testing, rather than as a universal static model.

\begin{figure}
    \centering
    \includegraphics[width=\linewidth]{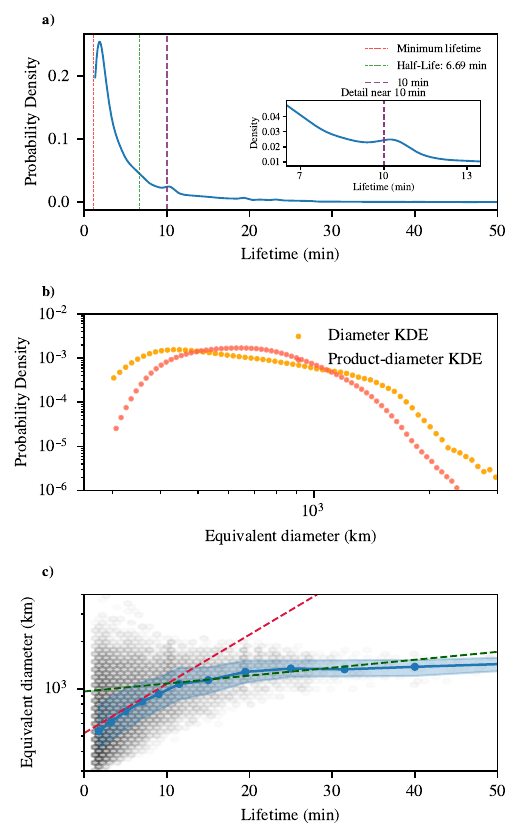}
    \caption{Physical properties of the tracked granules. \textbf{a)} Distribution of the lifetimes of the detected granules. The red dashed vertical line at $\simeq1$~min indicates the minimum lifetime imposed in the tracking procedure for a feature to be considered a valid detection. The green dashed line indicates the granule half-life, $\simeq7$~min, estimated from an exponential fit to the decay of the lifetime distribution. \textbf{b)} Distribution of the equivalent diameter of the detected structures (orange). The equivalent diameter is estimated under the assumption that the features are circular, which may result in an underestimation of the characteristic size of irregularly shaped granules. The red curve shows the distribution of the pairwise geometric-mean diameter, $\sqrt{D_iD_j}$, obtained by Monte Carlo sampling of $5\times10^6$ independently and uniformly drawn pairs. \textbf{c)} Relationship between feature lifetime and equivalent diameter. The grey hexagonal bins represent the logarithmic density of features, while the blue points indicate the conditional median equivalent diameter and the shaded region its interquartile range. The dashed red and green lines show the linear fits in $\log_{10}(D)$ to the short- and long-lifetime regimes, respectively, highlighting a transition at approximately $10$~min.}
    \label{fig:5}
\end{figure}

The superior performances of the UNET++ model can finally be visually assessed by the readers in Figure~\ref{fig:4}. Here, the qualitative differences between the intermediate and final steps of our approach can be clearly appreciated. The adaptive method alone struggles to capture the structural morphology of solar granulation, leading to artificial internal cavities and eroded borders. On the other hand, the UNET++ model predicts solid, contiguous granules that closely mirror the ground truth. 

\begin{figure}
    \centering
    \includegraphics[width=\linewidth]{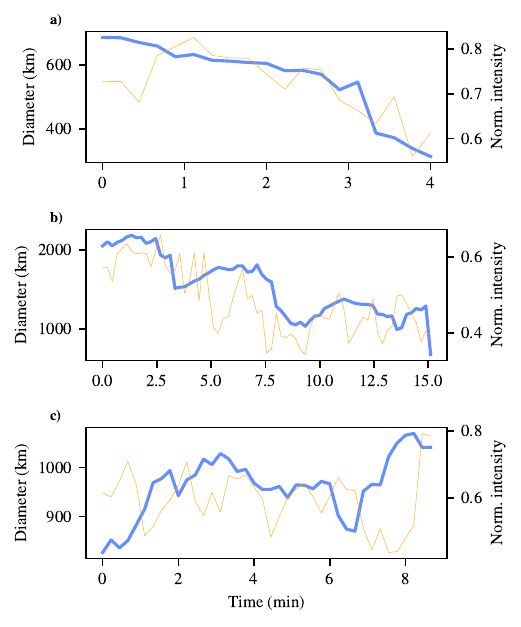}
    \caption{Sample of three different equivalent diameter (blue) and normalised intensity (yellow) time series extracted by three separate granules in our field of views. The strikingly different patterns observed in the curves highlight the hihgly variable dynamics of turbulent convection observed at such scales.}
    \label{fig:6}
\end{figure}

Fig.~\ref{fig:5} shows the distributions for the lifetimes and equivalent diameters of the tracked granules in the three different datasets with the Kernel Density Estimation (KDE). KDE applies Gaussian kernel smoothing to estimate the probability density function, providing a continuous representation of the distribution that avoids the binning artefacts inherent to standard histograms. Starting with the lifetime distribution at the top, as expected from solar granulation observed at this height, we see a mixture of different features with lifetimes ranging from about three to more than 20 minutes as an upper boundary. It is interesting to note the bump in the distribution at about 8 to 10 minutes, a timescale that is usually associated with the clustering timescale of stable and recurrent granules, as shown in \cite{berrilli_spatial_2005}. The half-life time was obtained by estimating the lambda decay time using an exponential fit and was found to be equal to $\simeq7$~minutes, which is compatible with references in the literature. Longer-lived tracks are likely to be granules that repeatedly reform in the same spot, making several successive cells look like a single persistent feature. This behaviour is later shown in panel b of Fig. \ref{fig:6}. In the middle panel of figure~\ref{fig:5}, we show the distribution of the equivalent diameter (ED) of the tracked granules (orange), which is consistent with expectations from simulations and previous studies of small-scale granular structures at very high spatial resolution \citep[see, e.g.,][]{2005A&A...429..335V, 2012ApJ...756L..27A}. However, the increased spatial resolution provided by DKIST allows us to extend the observed distribution towards smaller spatial scales. Interestingly, a bump analogous to that observed in the lifetime distribution is also present here, which we associate with the characteristic size of larger, barely resolved granules. The pairwise geometric-mean distribution, shown in red, provides an independent estimate of the characteristic spatial scale of the granular structures. Since it is constructed from the full population of pairwise combinations rather than from individual measurements, its peak is less sensitive to the scatter and noise that affect individual detections. It therefore converges towards the spatial scale around which the majority of granules in our datasets cluster, which is found to be approximately $\simeq700$--$800$~km. Moreover, under a purely self-similar hierarchy, in which parent structures fragment into daughter cells following a scale-free, power-law distribution, the pairwise geometric-mean distribution is expected to closely resemble the original ED distribution. The clear difference between the two distributions, therefore, indicates the presence of a preferred intrinsic spatial scale in the granular population. Finally, in the bottom panel of Fig.~\ref{fig:5}, we show the relationship between the lifetime and the equivalent diameter of the tracked granules. As previously reported by \cite{hirzberger_time_1999}, we identify a clear transition at approximately $10$ minutes, separating two distinct regimes in the evolution of the granular size, with a slope of, respectively, $0.031$ and $0.005$~dex/min, which corresponds to a growth rate of $7.5$\% and $1.2$\%. For lifetimes shorter than $\simeq10$ minutes, the equivalent diameter increases with lifetime, while for longer-lived granules it approaches a nearly constant value of $\simeq1500$~km. This change in behaviour suggests a transition between two distinct evolutionary regimes of solar granules, as also discussed by \cite{espagnet1993turbulent} and \cite{nordlund1997stellar}.

In conclusion, Fig.~\ref{fig:6} presents three examples of granular structures exhibiting different temporal behaviours. Panel a shows feature 10, whose equivalent diameter and lifetime are close to the peaks of the distributions shown in Fig.~\ref{fig:5}. Its morphology and temporal evolution suggest that it is a fragment produced by the splitting of a larger granular structure, subsequently shrinking and fading towards the end of its life. Panel b shows a particularly interesting case of a longer-lived structure evolving within a highly dynamic region of our field of view. The accompanying movie suggests an ongoing process of fragmentation, with portions repeatedly separating from the main granular body. Despite this repeated splitting, the remaining structure persists and undergoes further expansion, potentially sustained by the continued emergence of plasma. The interplay between expansion and fragmentation produces several distinct variations in equivalent diameter, accompanied by less pronounced variations in intensity. Finally, panel c illustrates the continuous interplay of merging and splitting within a group of granules, which ultimately coalesce into a single larger structure. This evolution highlights the highly dynamic nature of granulation at spatial scales below 1000~km. The dynamics discussed above are further illustrated by the videos provided in the supplemental materials.

\section{Conclusions}

In this work, we presented a novel hybrid segmentation technique for solar granules in the photosphere (SoFTg), combining the robustness and controllability of a classical adaptive-threshold approach with the speed and refinement capabilities of a deep-learning model such as UNET++. The results demonstrate that SoFTg can accurately identify and segment individual granules while retaining the flexibility and computational efficiency required for the analysis of large high-resolution datasets.

Applied to DKIST observations, SoFTg provides a robust framework for characterising granulation at spatial scales that are difficult to access with previous observational datasets. The resulting distributions of granule lifetimes and equivalent diameters show strong agreement with established observational results, providing an important validation of the method while demonstrating its applicability to the substantially higher spatial resolution of DKIST. At the same time, the high spatial resolution enables SoFTg to resolve and characterise smaller-scale structures that would be unresolved or less clearly identified in lower-resolution observations. This combination of consistency with established results and sensitivity to finer spatial scales makes SoFTg a valuable tool for systematic studies of solar granulation across a broad range of spatial scales.

The SoFTg code is freely available at \url{https://github.com/mib-unitn/SoFTg}.

\begin{acknowledgments}

MB acknowledges that this article was produced while attending the PhD program in  PhD in Space Science and Technology at the University of Trento, Cycle XXXIX, with the support of a scholarship financed by the Ministerial Decree no. 118 of 2nd March 2023, based on the NRRP - funded by the European Union - NextGenerationEU - Mission 4 ``Education and Research'', Component 1 ``Enhancement of the offer of educational services: from nurseries to universities'' - Investment 4.1 ``Extension of the number of research doctorates and innovative doctorates for public administration and cultural heritage'' - CUP E66E23000110001.

We wish to acknowledge the Space It Up project funded by the Italian Space Agency, ASI, and the Ministry of University and Research, MUR, under contract n. 2024-5-E.0 - CUP n. I53D24000060005.

MB gratefully acknowledges the financial support from the National Solar Observatory that made this publication possible.

We wish to acknowledge scientific discussions with the Waves in the Lower Solar Atmosphere (WaLSA; \href{https://WaLSA.team}{www.WaLSA.team}) team, which has been supported by the Research Council of Norway (project no. 262622), The Royal Society (award no. Hooke18b/SCTM; \citealt{2021RSPTA.37900169J}), and the International Space Science Institute (ISSI Team 502).

The research reported herein is based in part on data collected with the Daniel K. Inouye Solar Telescope (DKIST), a facility of the U.S. National Science Foundation. DKIST is operated by the National Solar Observatory under a cooperative agreement with the Association of Universities for Research in Astronomy, Inc. DKIST is located on land of spiritual and cultural significance to Native Hawaiian people. The use of this important site to further scientific knowledge is done with appreciation and respect.

\end{acknowledgments}

\begin{contribution}

All authors contributed equally to the presented work.

\end{contribution}

\facilities{VBI (DKIST)}

\software{PyTorch \citep{AnselPyTorch2024}, Astropy \citep{astropy:2013, astropy:2018, astropy:2022}}

\appendix

\section{Pre-processing comparison}

\begin{figure*}[ht!]
    \centering
    \includegraphics[width=\linewidth]{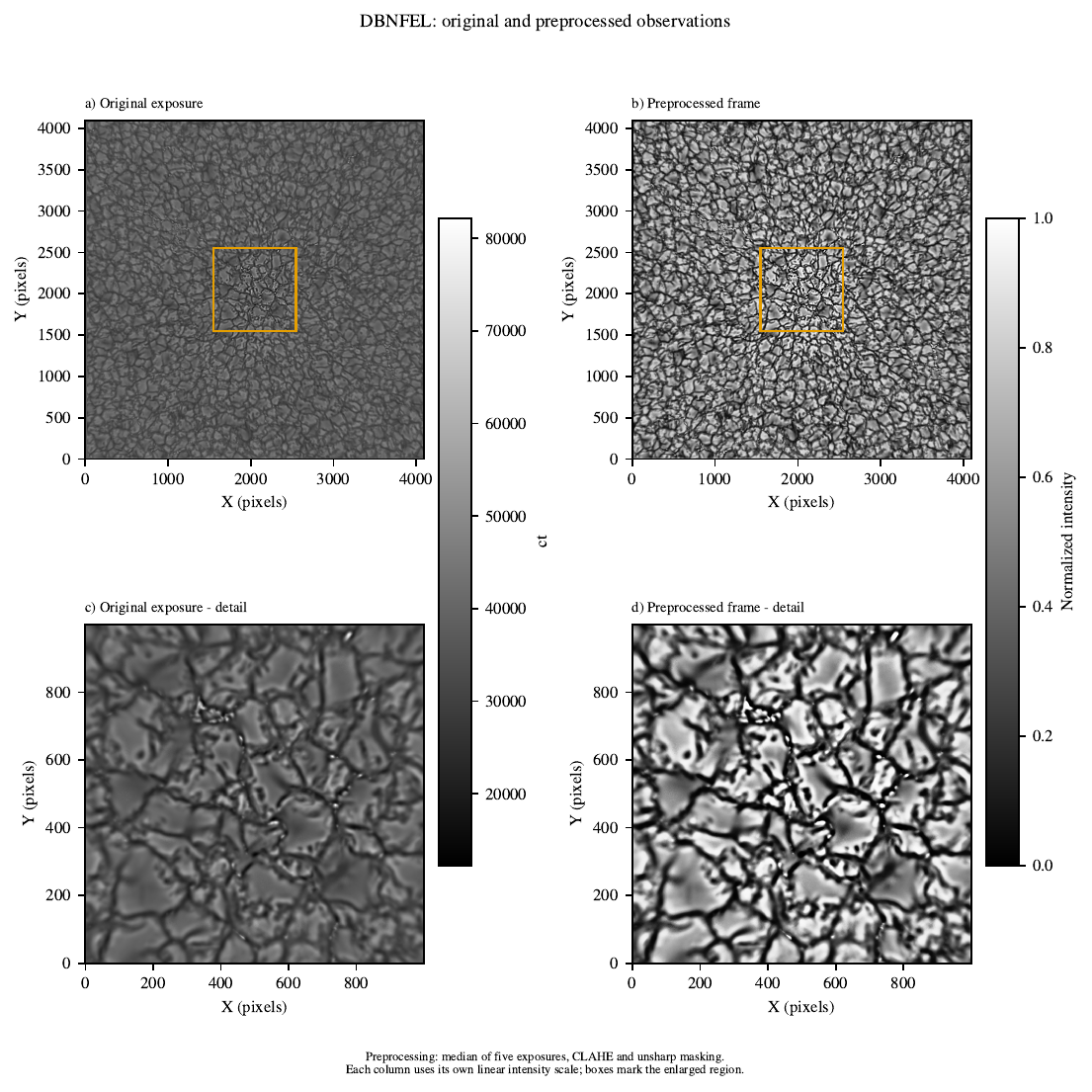}
    \caption{Comparison between the original speckle-reconstructed observation and the output of the pre-processing pipeline for a sample frame of the DBNFEL dataset. \textbf{a)} Full field of view ($4096 \times 4096\text{ pixels}$) of the original exposure, with the colour bar indicating raw intensity in detector counts ($\text{ct}$). \textbf{b)} The same field of view after applying the pre-processing pipeline (5-frame temporal median stacking, CLAHE, and unsharp masking), scaled to normalised intensity units $[0, 1]$. The orange squares in panels (a) and (b) outline the $1000 \times 1000\text{ pixel}$ central sub-field enlarged in the bottom row. \textbf{c)} Enlarged view of the original exposure. \textbf{d)} Enlarged view of the pre-processed frame, highlighting the sharpened granular boundaries, enhanced contrast of the dark intergranular lanes, and clearer resolution of internal granular substructures.}
    \label{fig:ap1}
\end{figure*}

Fig.~\ref{fig:ap1} illustrates the visual impact of the pre-processing pipeline described in Section~\ref{subsec:prepro} on a representative frame of the DBNFEL dataset. In the original speckle-reconstructed exposure (panel a), the full $4096 \times 4096\text{ pixel}$ field of view exhibits a relatively narrow dynamic range in raw detector counts ($\text{ct}$) and moderate contrast between bright granular interiors and dark intergranular lanes. After applying the five-frame temporal median stack, Contrast Limited Adaptive Histogram Equalisation (CLAHE), and unsharp masking (panel b), the intensity is normalised to the $[0, 1]$ range with uniform local contrast across the entire field of view. The advantages of this enhancement are particularly evident in the $1000 \times 1000\text{ pixel}$ enlarged central sub-fields (marked by the orange boxes in the top panels and shown in detail in panels c and d). While the original exposure (panel c) displays shallow intensity gradients across thin intergranular lanes, nascent splitting regions, and the darker cores typical of exploding granules, the pre-processed frame (panel d) sharply delineates the intergranular network and accentuates fine-scale internal substructures while suppressing pixel-scale noise. This produces steeper intensity transitions at the cell boundaries, providing an optimal input for both the adaptive thresholding step and the subsequent UNET++ segmentation.

\section{Augmentation process}

The data augmentation pipeline is divided into two steps, an ``offline'' pre-computation augmentation carried out when loading the dataset for the first time, and an ``online'' augmentation which is performed live when fetching each batch. 

\begin{figure}[htbp]
    \centering
    \includegraphics[width=\textwidth]{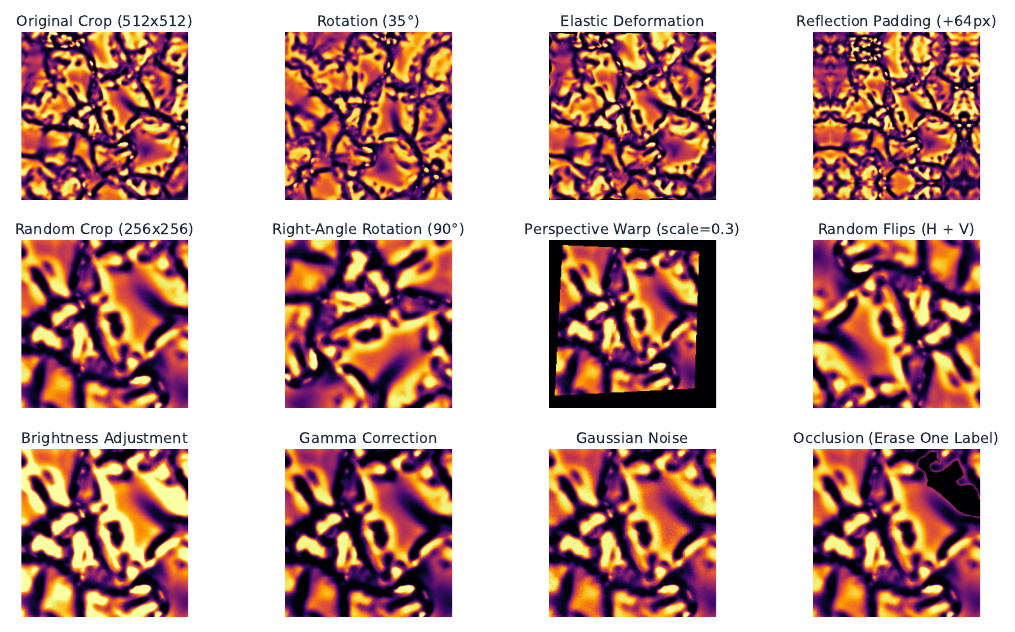}
    \caption{Visual representation of the offline and online data augmentation pipeline applied to the dataset.}
    \label{fig:augmentation_preview}
\end{figure}

Offline augmentation creates multiple variants of the same image with geometric distortions. To this extent, the applied modifications are as follows:

\begin{itemize}
    \item Rotation, performed at 18 distinct angles that ranged from $0^\circ$ to $85^\circ$ in steps of $5^\circ$ around the geometric centre of the padded image.
    \item Elastic deformation, implemented by synthesising a random spatial displacement field drawn from a normal distribution, smoothed with a Gaussian filter ($\sigma = 10$), and applied to both intensity images and class masks by spatial grid sampling.
    \item Reflection padding, which mirrors border pixels outward symmetrically to extend the canvas width based on the image diagonal, preventing edge clipping and boundary artefacts during arbitrary-angle rotations.
\end{itemize}

The dynamic augmentation is instead focused on spatial and photometric modifications to the already geometrically augmented images. These modifications are as follows:

\begin{itemize}
    \item Random crop, extracting a fixed-size $128 \times 128$ square window using a probability weighting distribution derived from a smoothed Gaussian kernel ($\sigma = 14$).
    \item Right-angle rotation, randomly chosen from $\{0^\circ, 90^\circ, 180^\circ, 270^\circ\}$.
    \item Random perspective with a deformation scale of 0.3 and a probability of $p = 0.5$ using the nearest-neighbour interpolation.
    \item Random flips, both horizontal and vertical, each applied independently with a probability of $p = 0.5$.
    \item Adjustment of brightness, multiplying the image by a factor uniformly sampled from $(0.8, 1.2)$ with a probability of $p = 0.7$.
    \item Gamma correction, increasing the clamped tensor to a power uniformly sampled from $(0.75, 1.35)$ with a probability of $p = 0.5$
    \item Gaussian noise, adding additive standard deviation noise sampled from $(0.005, 0.02)$ with a probability of $p = 0.5$.
    \item Occlusions, achieved by random erasing with a probability of $p = 0.3$, a scale range of $(0.02, 0.1)$, and an aspect ratio of $(0.3, 3.3)$.
\end{itemize}

All of the images are then normalised to constrain the intensity values between zero and one. The offline phase expands each raw input file into 18 augmented variants. Although the data set does not pre-calculate static global totals per class and instead dynamically produces batches of length $l = 1000$ per epoch, this multi-step process significantly expands the training variability across the two defined classes (Intergranular lanes and Granules). An example of the different augmentation transformations is shown in Figure~\ref{fig:augmentation_preview}.

\bibliography{main}
\bibliographystyle{aasjournal}

\end{document}